\documentstyle[aps,prl,twocolumn,epsfig]{revtex}

\begin{document}
\draft
\flushbottom
\twocolumn[
\hsize\textwidth\columnwidth\hsize\csname @twocolumnfalse\endcsname

\title{The Striped Phase in Presence of Disorder
and Lattice Potentials}
\author{N.~Hasselmann$^{a,b}$, 
A.~H.~Castro Neto$^a$, C.~Morais Smith$^{b}$, and Y.~Dimashko$^{b}$}
\address{
$^{a \,}$Dept. of Physics, University of California, Riverside, CA, 92521, 
USA \\
$^{b \,}$I. Institut f{\"u}r Theoretische Physik, Universit{\"a}t 
Hamburg, D-20355 Hamburg, Germany \\ 
}

\widetext

\date{\today}
\maketitle
\tightenlines
\widetext
\advance\leftskip by 57pt
\advance\rightskip by 57pt

\begin{abstract}
We study the influence of disorder and lattice effects on
the striped phase (SP) and the incommensurate (IC) spin fluctuations
of the cuprates and nickelates. Starting from
a phenomenological model of a discrete quantum string on a lattice
with disorder, we derive the renormalisation 
group (RG) equations in leading order in the lattice and 
disorder strength.
Three regimes are identified, the quantum membrane phase,
the flat phase pinned by the lattice and the disorder pinned 
phase. We compare our results with  
measurements on nickelates
and cuprates and find good agreement. 

\end{abstract}

]
\narrowtext

The discovery of a static SP 
in the compound $\rm La_{1.6-x}Nd_{0.4}Sr_xCuO_4$ at 
both non-superconducting
($x\simeq 1/8$) \cite{tranquada1} and superconducting 
($x=0.15, 0.20$) \cite{tranquada2}
compositions has revived
interest in the possibility that a similar, 
albeit fluctuating
order may be common to all under- and optimally doped
cuprate superconductors. 
The strongest indications for a fluctuating SP order exist for
the 
$\rm La_{2-x}Sr_xCuO_4$ family, where the occurrence of incommensurate (IC)
magnetic
peaks in inelastic neutron scattering experiments
has been observed for quite some time \cite{LaSrCuO,yamada,wells}.
Very recently, similar IC correlations were also measured in
$\rm YBa_2Cu_3O_{7-\delta}$ \cite{YBCO}. Also, evidence for stripes in
$\rm Bi_2Sr_2CaCu_2O_{8+\delta}$ has been reported \cite{bianconi}.
A static SP order has been  detected  in both O and
Sr doped nickelates 
($\rm La_{2-x}Sr_xNiO_{4+\delta}$) \cite{NiSr,NiO,O413},
which have a structure very similar to the cuprates. 
While the concept of a dynamical fluctuating SP 
accounts qualitatively for many properties
of cuprate superconductors (e.~g.~ the form of the
Fermi surface \cite{antonio1}, the IC spin fluctuations
\cite{zaanen1}, or the suppression of the N\'{e}el temperature with
doping \cite{antonio2}), it is 
at present theoretically not well understood. Even under the 
assumption that all excess holes are tightly bound in strings
many questions regarding the nature of the fluctuations, the charge- and
spin coherence of the ground state, and the influence of disorder or lattice 
potentials on the SP are still open.

In the long wavelength limit a sufficiently dilute array of classical 
stripes can be described
as a membrane with anisotropic stiffnesses, where the stiffness in one 
direction is dominated by the entropic repulsion between neighboring
stripes.
In presence of disorder, such a membrane is pinned at all temperatures,
leading to a glassy SP
\cite{nattermann}. 
Also, in the presence of a weak substrate lattice 
potential, a weakly
IC phase near a $p\times 1$ registered phase is known to be unstable to
the
formation of topological defects and hence a fluid if $p<\sqrt 8$
\cite{ictransition}. As
the hole stripes
in the cuprates act as domain walls to the antiferromagnetic (AF)
background, the SP
is topologically equivalent to an IC phase near $p=2$ which
led 
Zaanen {\em et al.} \cite{zaanen1} 
to the conclusion that the SP in the cuprates is
a stripe liquid.
However, the instability of the membrane
to both disorder pinning and formation of topological defects is rooted 
in the form of the entropic repulsion of classical strings. 
At low temperatures the fluctuations of the strings are
no longer thermal but predominantly quantum.
In this case
the effective steric interaction between
neighboring stripes decays exponentially with the distance 
rather than algebraically as in the purely thermal case \cite{zaanen1}. 
In this work we study the influence of disorder and lattice perturbations
in the quantum regime of the SP. We focus at the 
intermediate  time- 
and length scale regime, as e.g. probed in neutron scattering
experiments,
where the dynamics is dominated by a single stripe 
rather than the collective and coherent membrane physics. 
The neighboring stripes nonetheless play an important role as
they confine the stripe wandering. The stripe therefore
cannot find the optimal
path through the disorder potential. 

We start with a phenomenological
model of a stripe on a lattice in a disorder potential. 
Although our model
describes stripes oriented along one of the simple lattice
directions, the continuum limit we use below is not sensitive to
the microscopic details, so that our results also 
apply for a SP with a diagonal orientation. 
Implicitly we
assume that a particular SP order is  well
separated in energy from other configurations, so that through the RG the 
topology of the charge and spin order does not change.
To study the dynamics of the stripe, it is then possible to use a first
quantized 
formulation of the problem. Consider a directed string of
holes
(enumerated by  $n$) on a square lattice with lattice constant
$a$.
Each hole is only allowed to hop in the transversal direction. To account
for
the stripe stiffness, we include a parabolic potential of strength $K$
which
couples adjacent holes in the stripe. 
The Hamiltonian is then
$H=H_{S} + H_{D}$,
\begin{eqnarray}
H_{S}&=&\sum_n \left[-2 t \cos\left(\frac{\hat{p}_n a}{\hbar} \right) 
+ \frac{K}{2 a^2} 
\left( \hat{u}_n - \hat{u}_{n+1} \right)^2 \right] \ ,
\nonumber
\\
H_{D}&=&\sum_n V_n(\hat{u}_n) \ .
\nonumber
\end{eqnarray}
Here, $t$ is the hopping parameter and $\hat{p}_n$, $\hat{u}_n$ 
are canonical conjugate transversal momentum and 
position variables of the $n$-th hole, respectively. 
The eigenvalues of $\hat{p}_n$ 
are restricted to $-\pi < p_n a /\hbar < \pi $. 
The eigenvalues of $\hat{u}_n$ are thus integer
multiples of $a$. A model similar to $H_{S}$, without  disorder 
but with additional curvature terms, was investigated in
\cite{zaanen2}.
$H_{S}$ can be brought via the
canonical transformation $\hat{\varphi}_n=\sum_{m<n}\hat{p}_m$, 
$\hat{\pi}_n=\hat{u}_{n-1}-\hat{u}_n$ into a form previously studied
within the context of 1D Josephson Junction arrays \cite{bradley} and
exhibits  a $T=0$ Kosterlitz-Thouless (KT) \cite{kt} transition at a critical 
value of $t/K$. 
$H_{D}$ describes the interaction of the stripe with
an uncorrelated disorder potential,
$\left<V_n(u) V_{n'}(u') \right>_D = D \delta(u-u') \delta_{n,n'},$
where $\big< ... \big>_D$ denotes the gaussian average over the disorder
ensemble
and $D=\hbar^2 c/ \tau_I$. Here, $c$ is the characteristic velocity of
the stripe excitations and $\tau_I$ is the impurity scattering time.
Due to the stripe repulsion $|u|<L$, where
$L$ is the average inter-stripe distance.

The partition function $Z= tr \exp\left(-\beta H\right)$ can be
transformed
in the usual way into a path integral form. We introduce 
$\epsilon=\beta \hbar /M$ and insert $M-1$ times the resolution of unity
into
the trace. Using the Villain approximation $\exp{\left(
\ell \cos \phi \right)} \simeq \sum_q \exp{
\left(-q^2/(2 \ell) + i q  \phi \right)}$ and integrating over
the $p$'s and $q$'s we arrive at
\begin{eqnarray}
Z & \simeq & \sum_{ \{ u_{nm} \} } \exp \left\{ - \frac{1}{\hbar}
\sum_{m=0}^{M-1} \sum_{n}
\left[ 
\frac{\hbar^2}{4 t \epsilon a^2}\left( u_{nm+1}-u_{nm} \right)^2 
\right. \right. \nonumber \\ && \left. \left. 
+ \frac{\epsilon K}{2 a^2} \left( u_{nm}-u_{n+1,m} \right)^2 
+ \epsilon V(u_{nm})
\right] 
\right\} ,
\nonumber
\end{eqnarray}
where $m$ and $n$ are discrete indices for the  imaginary
time and longitudinal direction, respectively. After
introducing $i=1, \dots, N$ replicas 
and taking the continuum limit $\epsilon, a \to 0$, we find
the replica action
\begin{eqnarray}
\label{action}
S^r & = & \sum_i S_0[\phi^i] 
+ \frac{g}{a} \sum_i \int_0^{\hbar \beta} d\tau \int dy 
  \ \cos \left( 2 \sqrt{\pi} \phi^i \right)
\nonumber \\ & &
+ \frac{D}{2 a L\hbar} \sum_{i,i'} \int dy \int_0^{\hbar \beta} d\tau
d\tau'
\nonumber \\ & &
  \ \cos \left[ 2 \sqrt{\pi} \delta  
\left(\phi^i(y,\tau)-\phi^{i'}(y, \tau')\right) \right]
\end{eqnarray}
where $\delta=a/(2L)$ is essentially the stripe density. The
gaussian action $S_0$ is given by
\begin{eqnarray}
S_0[\phi(y,\tau)]= \frac{\hbar}{2 \pi \mu}
\int_0^{\hbar \beta} d\tau \int dy
\left[ \frac{1}{c}\left(\partial_\tau \phi \right)^2
+ c \left(\partial_y \phi \right)^2 \right].
\nonumber
\end{eqnarray}
Here, we introduced the dimensionless fields $\phi^i=\sqrt{\pi} u^i /a$.
The dynamics of the free stripe is completely characterized
by the velocity $c=a \sqrt{2 t K}/ \hbar$ and the dimensionless
parameter $\mu=\sqrt{2t/K}$, which is a measure for the competition between
kinetic and confining energies.
The parameter $g$ is introduced to account for the lattice effects.
In the derivation we
have used
$\delta(u)= (1/2 L) \sum_l e^{i \pi l u/L}$
and kept only the most relevant modes with $l=\pm 1$. 
The  form of the delta function reflects
the confinement of the stripe by its neighbors.

We now state more precisely under which conditions
the interaction between neighboring stripes can be neglected.
It is reasonable to assume
that the stripe-stripe interaction is dominated by contact interaction
resulting from a short range repulsion.
A simple estimate for the range
of validity of the single stripe description was proposed in \cite{zaanen1}:
equating
the mean square transversal wandering, 
$\Delta u(y, \tau)
\simeq \sqrt{\mu}\ a \ \pi^{-1} \ln \left[(y^2 + c^2 \tau^2)/a^2\right]$, 
with the 
average distance between stripes, $L$, immediately yields
$y_c = \tau_c c  \simeq a \exp [\pi/(4 \ \delta \sqrt{\mu})]$. 
Therefore, on time and length scales smaller than $y_c$, $\tau_c$, the
physics is dominated by single stripe dynamics while for larger scales
the interaction must be taken into account.
Although we cannot rule out the importance of
other possible sources
of direct stripe-stripe interactions, as e.g. (screened) Coulomb
forces or Casimir-type interaction \cite{pryadko},
at least for large $L$ we expect them to be weak
compared to the steric repulsion. However, should other forces 
dominate, the crossover would occur at scales
smaller than $y_c, \tau_c$.

To obtain the phase diagram of the stripe in presence of both disorder and
lattice potentials, we derive the RG differential equations
by calculating the correlator 
$\left<\exp i \sqrt{\pi} 
\left[ \phi^{i_1}(y_1, \tau_1) - \phi^{i_2}(y_2, \tau_2) \right] 
\right>$ 
to lowest non-vanishing order in $g$ and $D$.
Using the method
developed by Giamarchi and Schulz \cite{giamarchi1},
we find under the rescaling $a'=e^{\ell} a$ the flow equations  
\begin{eqnarray}
\begin{array}{lcl}
\displaystyle{\frac{d}{d \ell}} {\cal D}  
=  (3-2 \pi \mu \delta^2) {\cal D} \ , & &
\displaystyle{\frac{d}{d \ell}} {\cal G}  =  (2 - \pi \mu) {\cal G} \ ,\\  \\
\displaystyle{\frac{d}{d \ell}} \mu  =   
- \displaystyle{\frac{1}{2}}\mu^2 ({\cal G}^2 + {\cal D}) \ , & &
\displaystyle{\frac{d}{d \ell}} c  =  - \frac{1}{2} {\cal D} \mu c \ .
\end{array}
\label{rg}
\end{eqnarray}
where
${\cal D} =  4 \pi^2 D \delta^2 a^2/(\hbar^2 c^2 L)$ 
and ${\cal G} =  \pi^{3/2} g a/(c \hbar)$.
Our present analysis is strictly correct only for time- and
length scales smaller than $\tau_c$, $y_c$, at which the 
interaction between stripes becomes important. 
Nonetheless, these RG equations do strongly influence also the
zero frequency and wave vector behavior of the stripe array as we will
discuss below. We begin however the analysis ignoring these cutoffs.

With no disorder (${\cal D}=0$), the set of equations (\ref{rg}) 
reduces to the conventional KT form 
describing the roughening
transition of the stripe at $\mu_{c1}=2/\pi$.
For $\mu<\mu_{c1}$, ${\cal G}$ diverges,
signaling a pinning of the stripe by the underlying lattice.
The stripe is flat on length scales larger than the lattice
pinning length $L_{p1}\propto \exp[A/\sqrt{\mu_{c1}-\mu}]$ 
($A$ is a constant), 
has infinite stiffness
and its excitations are 
massive. For $\mu> \mu_{c1}$, the lattice potential is irrelevant and the
system flows towards a gaussian fixed point with renormalized $\mu^*$,
massless excitations, logarithmic wandering
and strong quantum fluctuations.

In the presence of disorder there is another instability at 
$\mu_{c2}=3/(2 \pi \delta^2)$ at which ${\cal D}$ becomes relevant. If
$\mu < \mu_{c2}$ (and ${\cal G}=0$), ${\cal D}$ flows to infinity and the
stripe is in a disordered (pinned) state. 
Near the transition it is possible to define a localization or collective
pinning length $L_{p2}$ of the stripe. Its asymptotic dependence
on the disorder strength near the critical
region (and ${\cal G}=0$) can be found from linearizing and integrating
the RG-equations close to the singularity,
$L_{p2}\propto \exp [B
\delta/\sqrt{({\cal D}_{0}-{\cal D}_{c})}]$ ($B$ is a constant),
where 
${\cal D}_{0}$ is the bare and ${\cal D}_{c}$ the critical disorder strength. 
Although the transverse excitations of the disorder pinned stripe 
are localized, the disorder does not lead to a gap in the density of
states \cite{mott}.
Further, contrary to the lattice pinned stripe, the stiffness of the
disorder pinned stripe is finite: adding a small linear tilt to the
fields $\phi^{i}\to \phi^{i}+\delta \phi \, y$ leaves the disorder term
in the action (\ref{action}) unchanged and the gaussian term in (\ref{action})
is modified only at  order  $(\delta \phi)^{2}$. Hence, 
the kink free energy vanishes \cite{mott}.

The diagram in Fig.~1 shows the stability regions 
of the stripe in the $\delta$-$\mu$ plane, for 
${\cal D}, {\cal G}\simeq 0$. If neither disorder nor lattice potentials
are relevant, the stripe is in the freely fluctuating gaussian phase.
On large scales, an array of coupled free gaussian strings can 
be described in terms of a 2+1D quantum membrane (QM).
If the lattice is relevant and disorder is not, the stripe is in
the flat phase.
The disordered stripe, which exists if disorder is relevant,
can be further distinguished
whether or not the lattice pinning is a relevant perturbation.
Even though disorder will always win in the long wavelength limit,
the stripe will be locally flat if the lattice potential is relevant
(``dis.~flat'') and if $L_{p1}<L_{p2}$. In the dis.~flat phase
the sharp band edge of the massive excitations which exist in the
flat phase are washed out by the disorder and Lifshitz-like tails
extend down to zero energy. No true gap survives but the 
lattice pinning reduces the low
energy spectral weight and gives rise to a pseudogap.
At $L_{p1}\approx L_{p2}$ the pseudogap  
disappears and a crossover to the disordered phase takes place.

\begin{figure}[v]
\epsfxsize=2.6 truein
\epsfysize=2.0 truein
\centerline{\epsffile{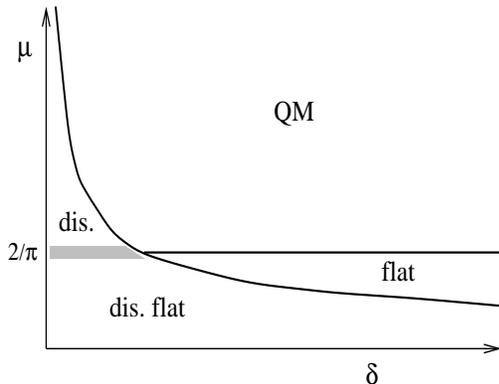}}
\caption[]{Phase diagram of the stripe as a function of $\delta=a/(2L)$ and
$\mu=\sqrt{2t/K}$.}
\label{pd}
\end{figure}

>From our analysis follows, that 
dynamical fluctuating SP order occurs only in a situation
in which both disorder and lattice potentials are irrelevant, implying
that the dynamical SP of the 
$\rm La_{2-x}Sr_{x}CuO_{4}$ compound must be insensitive to disorder.
Indeed, the IC spin fluctuations \cite{LaSrCuO}
in this compound near optimal doping
are strikingly similar to those found in
$\rm La_{2}CuO_{4+\delta}$ \cite{wells}. 
In $\rm La_{2}CuO_{4+\delta}$ 
the interstitial oxygens are mobile
down to temperature of about $200$ K and produce annealed as opposed
to the quenched disorder in the Sr doped compound. Although this
shows that weak disorder is unimportant near optimal doping, 
at lower doping, and hence lower $\delta$ 
(for $\rm La_{2-x}Sr_{x}CuO_{4}$, $\delta \propto x$ \cite{yamada}),
we expect from
Fig.~1 a critical doping $x_{c}$ at which the disorder becomes
relevant. Hence, below $x_{c}$ the stripes are pinned,
implying a broadening of the IC spin fluctuations.
Eventually, the IC peaks will overlap to
produce a single 
broad peak centered at the commensurate AF position, as has
been observed in the spin glass
phase ($x<0.05$) in neutron scattering 
\cite{yamada}. 
It is also likely that a new commensurate peak appears due
to the growth of strong AF fluctuations  and the appearance of phase-domain
walls close
to the AF ordered phase \cite{zeitz}. For  disorder pinned stripes,
a depinning transition under strong external fields has been predicted 
\cite{cris}.

In oxygen doped nickelates \cite{NiO,O413}, the
IC peak widths  are always much 
narrower than in the Sr doped ones \cite{NiSr} so
we conclude that disorder is relevant and hence $\mu<\mu_{c2}$ for 
the nickelates. 
Further, both Sr and O doped nickelates
show strong commensuration effects, i.e. the stripe spacing tends to
lock in at values commensurate with the lattice spacing \cite{NiSr,NiO,O413}. 
The width
of a commensurate plateau depends on the competition between the
effective stripe interaction and the strength of the lattice potential.
To compare these two effects, one must integrate over the independent
stripe fluctuations at smaller scales up to $y_c$. If $\mu<\mu_{c1}$,
the effective lattice coupling is strong and very wide commensurate
plateaus as a function of either doping or temperature are expected. This
is the situation in the nickelates. Similar commensuration effects 
are also observable in the cuprates, however, because
$\mu>\mu_{c1}$, the renormalized lattice coupling at $y_c$ is exponentially
suppressed and only  very narrow commensurate plateau exist. The 
SP is then nearly always in a floating phase IC with the lattice. 
This is consistent with the very weak Bragg peaks which are observable in
$\rm La_{2-x}Sr_{x}CuO_{4}$ only at $x=1/8$ \cite{suzuki} and the
static SP of  
$\rm La_{1.6-x}Nd_{0.4}Sr_xCuO_4$ which is most pronounced again at 
$x\simeq1/8$ \cite{tranquada1} whereas at other compositions the 
IC Bragg peaks are much weaker and broader \cite{tranquada2}.

\begin{figure}[v]
\epsfxsize=3 truein
\epsfysize=2 truein
\centerline{\epsffile{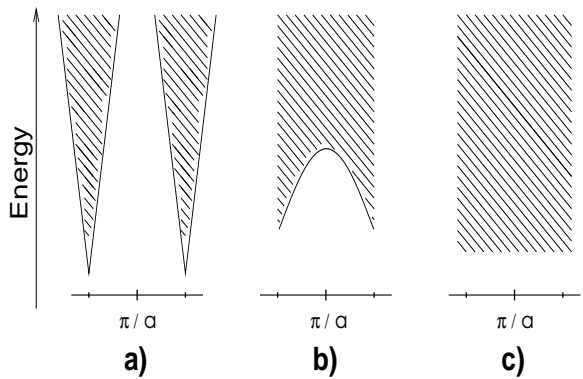}}
\caption[]{
Evolution of the IC signal with energy: a) for commensurate lattice pinned 
stripes and long range AF order; b) in the QM phase; c) for a strongly disordered SP.}
\label{ic}
\end{figure}

We now briefly discuss how the stripe fluctuations affect the correlations
of the spin system, where we now consider an array of stripes
with a harmonic coupling of strength $U$ between neighboring stripes. 
We first write the staggered spin density as 
${\bf M}({\bf r}, \tau)={\bf M}_{AF}({\bf r},\tau)
M_S({\bf r, \tau})$ 
where ${\bf M}_{AF}$ describes the staggered spin density of the
confined undoped regions and $M_S$ is a function which changes sign
at the position of the domain walls. 
It can be shown
that the inelastic part of the $\left<M_{S}M_{S}\right>$ correlator 
can be written at $T=0$ as \cite{next}
\begin{eqnarray} \label{fourier}
 \left<M_{S}M_{S}\right>_{\em inel}({\bf k},\omega)
 \propto  \Gamma(k_{x}\pm \pi/L, k_{y},\omega) 
e^{-k_{x}^{2} \left[\xi^{2}+  a^{2} \Gamma_{0}/\pi
\right] } && \ , 
\nonumber
\\ \nonumber \\ 
\Gamma({\bf k}, \omega)  
 \propto  [\omega^{2}+\omega_{k_{y}}^{2}+
2 U c^{2} \mu (1-\cos k_{x}L)/a^2]^{-1}
\nonumber &&
\end{eqnarray}
where $k_{x}$ ($k_{y}$) is the wave vector perpendicular (parallel) to the 
stripe orientation, $\xi$ the width of the stripes,
$\Gamma_{0}\simeq \pi \mu^{3/4}/(4U^{1/4})$  the quantum analog of a
Debye-Waller factor, $\Gamma$
the propagator of the stripe array and $\omega_{k_y}$ the single
stripe dispersion. From the form of the Debye Waller factor it is
seen that weak coupling (small $U$) and strong quantum fluctuations (large
$\mu$) suppress the inelastic signal.  
Assuming, that the dynamics of $\left<M_{S}M_{S}\right>$
and $\left< {\bf M}_{AF} {\bf M}_{AF} \right>$ decouple, the
$\left<{\bf M}{\bf M}\right>$ correlator becomes a simple convolution of
$\left<{\bf M}_{AF}{\bf M}_{AF}\right>$ and $\left<M_S M_S\right>$ in 
$({\bf k}, \omega)$-space. 
For commensurately pinned stripes
the stripe fluctuations are suppressed, but long ranged AF order 
can give rise to dispersion
at low energies through spin waves and an energy evolution as
drawn qualitatively in Fig.~\ref{ic}a. Such an energy dependence of
the peak width has been measured in $\rm La_2NiO_{4.133}$ \cite{O413}. 
If both lattice and disorder are irrelevant,
$\omega_{k_y}=c^* k_{y}$ and the stripe propagator supports both
acoustic modes near ${\bf k}=0$ and optical modes at $k_x=\pm\pi/L$.
In the cuprates, the AF order is short ranged which leads to a finite
energy broad peak around ${\bf k}=(\pm \pi/a, \pm\pi/a)$ of
$\left<{\bf M}_{AF}{\bf M}_{AF}\right>$. 
The evolution of the IC signal
has qualitatively the form as shown in Fig.~\ref{ic}b, with a dispersion
(due to stripe excitations)
away from the IC position to the commensurate position at
larger energies. The dispersion 
is not symmetric around the IC positions because the Debye Waller factor
suppresses signals far from the AF wave vector. 
This simple picture actually describes the experimentally
observed evolution of the IC peaks in 
$\rm La_{2-x}Sr_{x}CuO_{4}$ \cite{hayden}.
In the disorder
dominated SP, localized single stripe excitations 
lead to a broad spectral weight distribution in $\bf k$ space
at low energies (Fig.~\ref{ic}c) resulting in a broad featureless
peak centered at $(\pi/a,\pi/a)$, as is observed in the
spin glass phase of the cuprates \cite{yamada}.

In conclusion, using a perturbative RG calculation we  
studied the
competition between disorder and lattice effects on an effective model for
transverse stripe fluctuations. 
We identified three different phases, depending on the stripe density
$\delta$, 
the ratio between the
kinetic to the confining energy of the stripes ($\mu$), 
the strength of disorder ($\cal{D}$) and
the lattice strength (${\cal G}$): 
pinned by disorder, pinned
by the lattice and the QM phase. 
Given the experimental data
in cuprates and nickelates we can locate these materials
in our phase diagram and predict crossovers and
phase transitions between the distinct phases. Using a simple
model calculation, we discussed the energy evolution
of the IC spin fluctuations and obtained qualitative agreement with
the experiments.

We thank  A.~O.~Caldeira,
G.~Castilla, M.~P.~A.~Fisher,
A.~van Otterlo and H.~Schmidt for helpful discussions.
N.~H.~acknowledges support from the 
Gottlieb Daimler- und Karl Benz-Stiftung and the Graduiertenkolleg
``Physik nanostrukturierter Festk\"orper'', Univ.~Hamburg. 
A.~H.~C.~N. acknowledges support from the Alfred P.~Sloan foundation and
the US Department of Energy.

\end{document}